 \documentclass[12pt,preprint]{aastex}

\newcommand{\etal}{\mbox{et~al.}}

\def\deg      {{\ifmmode^\circ\else$^\circ$\fi}} 

 \shorttitle{COSMOS-HST}
 \shortauthors{Scoville et al.}
 
\begin{document}
 
 
 \title{COSMOS : Hubble Space Telescope Observations}
 
 
 \author{N. Scoville\altaffilmark{1,2}, 
R. G. Abraham\altaffilmark{3},
H. Aussel\altaffilmark{4,20},
J. E. Barnes\altaffilmark{4},
A. Benson\altaffilmark{1},
A. W. Blain\altaffilmark{1},
D. Calzetti\altaffilmark{5},
A. Comastri\altaffilmark{32},
P. Capak\altaffilmark{1},
C. Carilli\altaffilmark{6},
J. E. Carlstrom\altaffilmark{7},
C. M. Carollo\altaffilmark{8},
J. Colbert\altaffilmark{31},
E. Daddi\altaffilmark{9},
R. S. Ellis\altaffilmark{1},
M. Elvis\altaffilmark{10},
S. P. Ewald\altaffilmark{1},
M. Fall\altaffilmark{5},
A. Franceschini\altaffilmark{35},
M. Giavalisco\altaffilmark{5},
W. Green\altaffilmark{1},
R. E. Griffiths\altaffilmark{11},
L. Guzzo\altaffilmark{12},
G. Hasinger\altaffilmark{13},
C. Impey\altaffilmark{14},
J-P. Kneib\altaffilmark{15},
J. Koda\altaffilmark{1},
A. Koekemoer\altaffilmark{5},
O. Lefevre\altaffilmark{15},
S. Lilly\altaffilmark{8},
C. T. Liu\altaffilmark{33},
H. J. McCracken\altaffilmark{17,34},
R. Massey\altaffilmark{1},
Y. Mellier\altaffilmark{17},
S. Miyazaki\altaffilmark{18},
B. Mobasher\altaffilmark{5},
J. Mould\altaffilmark{9},
C. Norman\altaffilmark{19},
A. Refregier\altaffilmark{20},
A. Renzini\altaffilmark{21,35},
J. Rhodes\altaffilmark{1,22},
M. Rich\altaffilmark{23},
D. B. Sanders\altaffilmark{4},
D. Schiminovich\altaffilmark{24},
E. Schinnerer\altaffilmark{25},
M. Scodeggio\altaffilmark{38},
K. Sheth\altaffilmark{1,31},
P. L. Shopbell\altaffilmark{1},
Y. Taniguchi\altaffilmark{26},
N. D. Tyson\altaffilmark{16},
C. M. Urry\altaffilmark{27},
L. Van Waerbeke\altaffilmark{28},
P. Vettolani\altaffilmark{29},
S. D. M. White\altaffilmark{30},
L. Yan\altaffilmark{31},
G. Zamorani\altaffilmark{29}}

 
 
\altaffiltext{$\star$}{Based on observations with the NASA/ESA {\em
Hubble Space Telescope}, obtained at the Space Telescope Science
Institute, which is operated by AURA Inc, under NASA contract NAS
5-26555.}  
\altaffiltext{1}{California Institute of Technology, MC 105-24, 1200 East
California Boulevard, Pasadena, CA 91125}
\altaffiltext{2}{Visiting Astronomer, Univ. Hawaii, 2680 Woodlawn Dr., Honolulu, HI, 96822}
\altaffiltext{3}{Department of Astronomy and Astrophysics, University of Toronto, 60 St. George Street, Room 1403, Toronto, ON M5S 3H8, Canada}
\altaffiltext{4}{Institute for Astronomy, 2680 Woodlawn Dr., University of Hawaii, Honolulu, Hawaii, 96822}
\altaffiltext{5}{Space Telescope Science Institute, 3700 San Martin
Drive, Baltimore, MD 21218}
\altaffiltext{6}{National Radio Astronomy Observatory, P.O. Box 0, Socorro, NM
87801-0387}
\altaffiltext{7}{Department of Physics, University of Chicago, 5640 South Ellis Avenue, Chicago, IL 60637}
\altaffiltext{8}{Department of Physics, ETH Zurich, CH-8093 Zurich, Switzerland}
\altaffiltext{9}{National Optical Astronomy Observatory, P.O. Box 26732, Tucson, AZ 85726}
\altaffiltext{10}{Harvard-Smithsonian Center for Astrophysics, 60 Garden Street, Cambridge, MA 02138}
\altaffiltext{11}{Department of Physics, Carnegie Mellon University, 5000 Forbes Avenue, Pittsburgh, PA 15213}
\altaffiltext{12}{Osservatorio Astronomico di Brera, via Brera, Milan, Italy}
\altaffiltext{13}{Max Planck Institut f\"ur Extraterrestrische Physik,  D-85478 Garching, Germany}
\altaffiltext{14}{Steward Observatory, University of Arizona, 933 North Cherry Avenue, Tucson, AZ 85721}
\altaffiltext{15}{Laboratoire d'Astrophysique de Marseille, BP 8, Traverse
du Siphon, 13376 Marseille Cedex 12, France}
\altaffiltext{16}{American Museum of Natural History, Central Park West at 79th Street, New York, NY  10024}
\altaffiltext{17}{Institut d'Astrophysique de Paris, UMR7095 CNRS, Universit\`e Pierre et Marie Curie, 98 bis Boulevard Arago, 75014 Paris, France}
\altaffiltext{18}{Subaru Telescope, National Astronomical Observatory of Japan, 650 North Aohoku Place, Hilo, HI 96720.}
\altaffiltext{19}{Department of Physics and Astronomy, Johns Hopkins University, Homewood Campus, Baltimore, MD 21218}
\altaffiltext{20}{Service d'Astrophysique, CEA/Saclay, 91191 Gif-sur-Yvette, France}
\altaffiltext{21}{European Southern Observatory,
Karl-Schwarzschild-Str. 2, D-85748 Garching, Germany}
\altaffiltext{22}{Jet Propulsion Laboratory, Pasadena, CA 91109}
\altaffiltext{23}{Department of Physics and Astronomy, University of
California, Los Angeles, CA 90095}
\altaffiltext{24}{Department of Astronomy, Columbia University, MC2457,
550 W. 120 St. New York, NY 10027}
\altaffiltext{25}{Max Planck Institut f\"ur Astronomie, K\"onigstuhl 17, Heidelberg, D-69117, Germany}
\altaffiltext{26}{Physics Department, Graduate School of Science, Ehime University, 2-5 Bunkyou, Matuyama, 790-8577, Japan}
%
\altaffiltext{27}{Department of Astronomy, Yale University, P.O. Box 208101, New Haven, CT 06520-8101}
\altaffiltext{28}{Institut d'Astrophysique de Paris, 98 bis, boulevard Arago, F-75014 Paris, France.}
\altaffiltext{29}{INAF-Osservatorio Astronomico di Bologna, via Ranzani 1, I-40127 Bologna, Italy}
\altaffiltext{30}{Max-Planck-Institut f\"ur Astrophysik, D-85748 Garching bei M\"unchen, Germany}
\altaffiltext{31}{Spitzer Science Center, California Institute of Technology, Pasadena, CA 91125}

\altaffiltext{32}{INAF-Osservatorio Astronomico di Bologna, via Ranzani 1, 40127 Bologna, Italy}

\altaffiltext{33}{Astrophysical Observatory, City University of New York, College of Staten Island, 2800 Victory Blvd, Staten Island, NY  10314}
\altaffiltext{34}{Observatoire de Paris, LERMA, 61 Avenue de l'Observatoire, 75014 Paris, France}
\altaffiltext{35}{Dipartimento di Astronomia, Universitˆ di Padova, vicolo dell'Osservatorio 2, I-35122 Padua, Italy}
\altaffiltext{36}{Instituto di Astrofisica Spaziale e Fisica Cosmica, CNR, via Bassini 15, 20133 Milano, Italy}
%
%
  
 \begin{abstract}
 The Cosmic Evolution Survey (COSMOS) was initiated with an extensive allocation
 (590 orbits in Cycles 12-13) using the Hubble Space Telescope (HST) for high resolution imaging. Here we review 
 the characteristics of the HST imaging with the Advanced Camera for Surveys 
 (ACS) and parallel observations with NICMOS and WFPC2. A square field (1.8$\sq$\deg) has
 been imaged with single-orbit ACS I-F814W exposures with 50\% completeness for sources 0.5\arcsec~ in diameter at I$_{AB} $ = 26.0 mag. The ACS imaging is a key part of the COSMOS survey, providing very
 high sensitivity and high resolution (0.09\arcsec ~FWHM, 0.05\arcsec ~pixels) imaging and 
 detecting 1.2 million objects to a limiting magnitude of 26.5 (AB). These images yield resolved 
 morphologies for several hundred thousand galaxies. The small HST PSF also provides 
 greatly enhanced sensitivity for weak lensing investigations of the dark matter 
 distribution.
 \end{abstract}
 
 
 

 
 \section{Introduction} 
 
 Sensitive, high resolution imaging is a critical component of all cosmological 
 evolution studies, especially for surveys probing the evolution of luminous galaxies
 at redshift z $> 0.5$, when most galaxy assembly and evolution occurred. This
 approach was initiated in HST Treasury surveys : first, the HDFs  \citep{wil96,wil00} which imaged a 5 \sq\arcmin~
 area, followed by GOODS \citep{gia04} which covered a larger 
 area (360 \sq\arcmin), GEMS \citep{rix04} which was still more extensive (800 \sq\arcmin) but at shallower depth and, 
 most recently, the UDF survey  \citep{bec06} which was extremely deep but covered only 11 \sq\arcmin. 
The Cosmic Evolution Survey (COSMOS) with a 2 \sq\deg field, 
 is the first HST survey specifically designed to thoroughly probe the evolution of 
 galaxies, AGN and dark matter in the context of their cosmic 
 environment (i.e. large scale structure -- LSS). COSMOS 
 samples all relevant scales of LSS -- up to 
 $\sim$ 50 -- 100 h$_{70}^{-1}$ Mpc at all z $> 0.5$. The area of COSMOS was designed 
 to sample the full dynamic range of large scale structures from voids to very massive clusters. 
 (HST-ACS coverage of the DEEP Groth strip covers $\sim10\times70$\arcmin ~\cite{fab06},
 similar to GEMS, but the elongated geometry is not optimum for sampling the 
 larger structures.)
  High resolution imaging with HST enables 
 accurate determination of galaxy morphologies and multiplicities. The HST imaging 
 also provides significantly improved weak lensing analysis to probe the dark matter 
 distribution of the LSS.

COSMOS is the largest HST survey ever 
undertaken -- imaging an equatorial field 
with single-orbit I-band exposures to a depth I$_{AB} = \sim28$ mag (5$\sigma$ on an optimally extracted point source) and 50\% completeness for sources 0.5\arcsec~ in diameter at I$_{AB} $ = 26.0 mag.
With this area coverage, 
COSMOS HST and ground-based imaging detects $\simeq2\times10^6$ objects and samples a volume in the 
high redshift universe approaching that sampled locally by the Sloan Digital Sky Survey (SDSS).
In this article we describe the observations which make up the HST component 
of the COSMOS survey: the primary ACS imaging and the parallel NICMOS 
and WFPC2 imaging. A detailed description of the HST data processing 
is provided in \cite{koe06}. 

\section{HST Observations}

The original HST Cycle 12 COSMOS proposal had two major components for 
imaging with the Advanced Camera for Surveys (ACS) : 1) a complete mosaic of 2$\sq$\deg ~in the F814W 
(I-band) filter for morphological information and 2)
a similar mosaic in F475W (g-band) to provide resolved color imaging for studies of 
stellar populations and dust obscuration. Although 640 orbits were allocated for 
the I-band imaging, 50 orbits were specified to enable searches for SNIa. Following
discussions with the principal investigators for the supernova 
programs (A. Reiss \& S. Perlmutter), the COSMOS team decided that the 
SNIa science was not easily served within the COSMOS survey strategy,
and the 50 orbits were given over to follow-on exposures in the GOODS-S survey 
field. Thus the COSMOS HST survey (590 orbits total) comprises 
270 and 320 orbits allocated within HST Cycles 12 \& 13 (Fall 2003 to Spring 2005), 
respectively.

\subsection{Field Selection}

Multi-wavelength imaging and optical spectroscopy are central to the ability of the COSMOS survey 
to probe the evolution of stellar populations, star formation, galaxies and AGN. 
The enormous investments in observing time, required to cover a 2$\sq$\deg ~field, rule against
having separate northern and southern hemisphere fields, as in 
earlier, smaller surveys. Thus an initial prerequisite for COSMOS was that the 
field be accessible to telescopes in both hemispheres and especially all unique
facilities. This precluded using COSMOS to extend the area of earlier survey 
fields at high northern or southern Declination (such as HDF-N/S, 
GOODS/CDF-N/S, Lockman Hole and the Groth strip). 
An equatorial field is required to enable access by all existing 8-10m optical 
telescopes (and future larger telescopes) and the unique 
radio facilities (the (E)VLA in the north and ALMA in the south). ALMA is likely to
become a 'required' facility for studies of early universe galaxy 
evolution; at the same time, high sensitivity VLA radio imaging 
is critical and the instrument is unique in terms of sensitivity (with a factor of 3-10 improvement for
EVLA).

The original field proposed for the COSMOS survey was the VVDS/XMM Deep field centered 
at RA = 2$^{\rm h}$:26$^{\rm m}$, DEC = $-4.5$\deg which was scheduled for 
extensive optical spectroscopy with the VLT-VIMOS spectrograph. However, during the Phase 2 preparation for COSMOS-HST in Cycle 12
it became apparent that HST was very overcommitted at RA $\sim 2$ hr due to 
the Ultra-Deep Field (UDF) Directors Discretionary Time project in Cycle 12 -- we were therefore requested 
by both the STScI and ESO directors to consider shifting the COSMOS survey field to 
a non-conflicting RA. Our examination of alternative equatorial fields, revealed 
a field near the VVDS 10 hr field which in fact had slightly lower extinction 
(E$_{B-V} \simeq$  0.02 mag) and far-infrared, 
 cirrus backgrounds \citep{sco06a} than the original 2 hr VVDS/XMM Deep field. 
 This field also has no extremely bright X-ray, UV, optical and radio sources, unlike
 some other fields. 
 Since the 10 hr field did not have as extensive prior observational coverage 
 as the 2 hr field, most of the ground-based optical/ir imaging would have to be done 
 as part of COSMOS. However, this enables the COSMOS survey to efficiently
 observe the entire field to uniform sensitivity and to the optimum depth determined by 
 the COSMOS science.  
 
In conclusion,  the  final field selected for COSMOS is a 
1.4\deg$\times$1.4\deg ~square, aligned E-W, N-S, 
centered at RA = 10$^{h}$:00$^m$:28.6$^s$ , DEC = +02\deg:12\arcmin:21.0\arcsec (J2000). 
 It has {\it low and exceptionally uniform 
optical extinction}\citep[less than 20\% variation, see][]{san06},. Being equatorial, the field has somewhat higher far infrared 
background than the very best fields such as Lockman Hole \citep{sco06a}; however, 
 this is clearly less detrimental to 
the overall survey than the penalty of very poor (or nonexistent) radio and mm/submm 
coverage of higher declination fields by the VLA or ALMA. 

A summary of the HST observations is provided in Table \ref{tbl-1}; below, we
briefly discuss each instrument in detail.

\subsection{ACS Observations and Processing}

Imaging with ACS in the F814W (I-band) filter is the primary COSMOS HST observation. 
The ACS-WFC field of view (FOV) is 203\arcsec$\times203$\arcsec, covered by two CCD arrays separated by 
a gap of 4.5\arcsec. The pixel size is 0.05\arcsec. 
Nine of the allocated orbits were devoted to a test 3$\times$3 pointing mosaic in the 
F475W (g-band) filter at the center of the field in order to evaluate the need for 
full field coverage in a second filter. Thus a total of 581 orbits/pointings were devoted 
to imaging in the I-band filter. Within each orbit, four equal length exposures of 
507 sec duration each (2028 sec total) were obtained in a 4 position dither pattern, designed 
to shift bad pixels and to fill in the 90 pixel gap between the two ACS CCD arrays \citep[see ][]{koe06}. Adjacent pointings in the mosaic were positioned with approximately 4\% overlap in order to provide at least 
3 exposure coverage at the edge of each pointing and 4 exposure coverage over approximately
95\% of the survey area. This multiple exposure coverage with ACS provides excellent cosmic ray rejection \citep{koe06}. 

The visibility windows for the COSMOS program were set such that
two approximately 180\deg -opposed orientation angles could be scheduled 
(PA = 290$\pm$10\deg and 110$\pm$10\deg ~corresponding to 13 October to 7 January 
and 2 March to 21 May, respectively). (In the HST 2-gyro mode, only 
the former is available.) Three of the pointings had large 
reflections or scattered light due to bright stars being on the edge of the 
ACS FOV. These three fields were later repeated with two exposures placing the 
bright stars well {\it within} the ACS FOV. (They are included in the 581 orbit count
mentioned above.)

A full description of the ACS 
data processing including drizzling, flux calibration, registration and mosaicing 
is provided in  \cite{koe06}. The registration was tied into ground based 
CFHT i-band imaging \citep{aus06} with the USNO-B 1.0 reference frame offsets established 
from the COSMOS VLA survey \citep{sch06}. The absolute registration of all ACS data in the COSMOS 
archive is accurate to approximately 0.05-0.10\arcsec ~over the entire field \citep{koe06}.
The flux calibration of the ACS is tied into the standard STScI ACS calibration, 
accurate to better than 0.05\% in absolute zero point. For the public-released images with 0.05\arcsec ~pixels, the DRIZZLE parameters \citep{fru02} were : pixfrac = 0.8 and a square kernel was used. For the images used for weak lensing analysis, 0.03\arcsec ~pixels, pixfrac = 0.8, were used with a Gaussian kernel. (CTE effects 
on the faint source PSFs were reduced as described in \cite{rho06}.)
The final 
ACS mosaic image released to the public IRSA and MAST archives 
is sampled with 0.05\arcsec ~pixels. The measured FWHM of the PSF 
in the ACS I-band filter is 0.09\arcsec. These individual images were also 
rotated to North up for the public release data. For the purpose of the 
weak lensing analysis done by the COSMOS team an internal release 
of the unrotated images, sampled to 0.03\arcsec ~pixels, was also generated
to avoid rotating the original PSF and to reduce aliasing problems 
associated with resampling \citep{rho06}; for general morphological studies
the rotated and the 0.05\arcsec are entirely adequate.

\subsection{NICMOS and WFPC2 Parallel Observations}

In parallel with the ACS observations, imaging was obtained with 
 the NICMOS and WFPC2 cameras. WFPC2 was used in coordinated-parallel mode,
 implying that exposures were obtained with every ACS pointing. The WFPC2 
 FOV is 150\arcsec,  offset 
 in position from the ACS field center by 5.8\arcmin. Since NICMOS 
 cannot be used in periods of high particle flux such as the SAA passages, 
 the NICMOS observations were set up in pure-parallel mode so as 
 not to impede scheduling of the primary ACS observations. Therefore,
 not every orbit has an associated NICMOS parallel observation. The NICMOS Camera 
 3, used for COSMOS parallel observations, has a FOV of 50\arcsec ~and the field 
 center is displaced 8.5\arcmin ~from that of ACS. 
 
 For WFPC2, the filter used initially was F300W. However inspection of the Cycle 12 data
 revealed a very low rate of object detection at 3000\AA; in the second half of Cycle 13, 
 the WFPC2 filter was therefore changed to F450W. The NICMOS parallels used the 
 F160W (1.6$\mu$m, H-band) filter. There were 4 and 3 exposures per orbit for WFPC2 
 and NICMOS, respectively. The WFPC2 
 and NICMOS parallels cover approximately 55\% (1.07 \sq\deg) and 6\% (0.092 \sq\deg, 330 \sq\arcmin) of the COSMOS 1.8\sq\deg field
 imaged with ACS. The total areal coverage in NICMOS parallel 
 imaging is probably the largest of any HST project -- although it is not contiguous, it 
 provides enormous samples of objects. The details of 
 reduction and calibration of the WFPC2 (done by S. Ewald) and 
 NICMOS (done by J. Colbert) imaging are included in \cite{koe06}.
 
 \section  {Sensitivity, Resolution and Coverage}
 
\begin{figure}[ht]
\epsscale{1.}  
\plotone{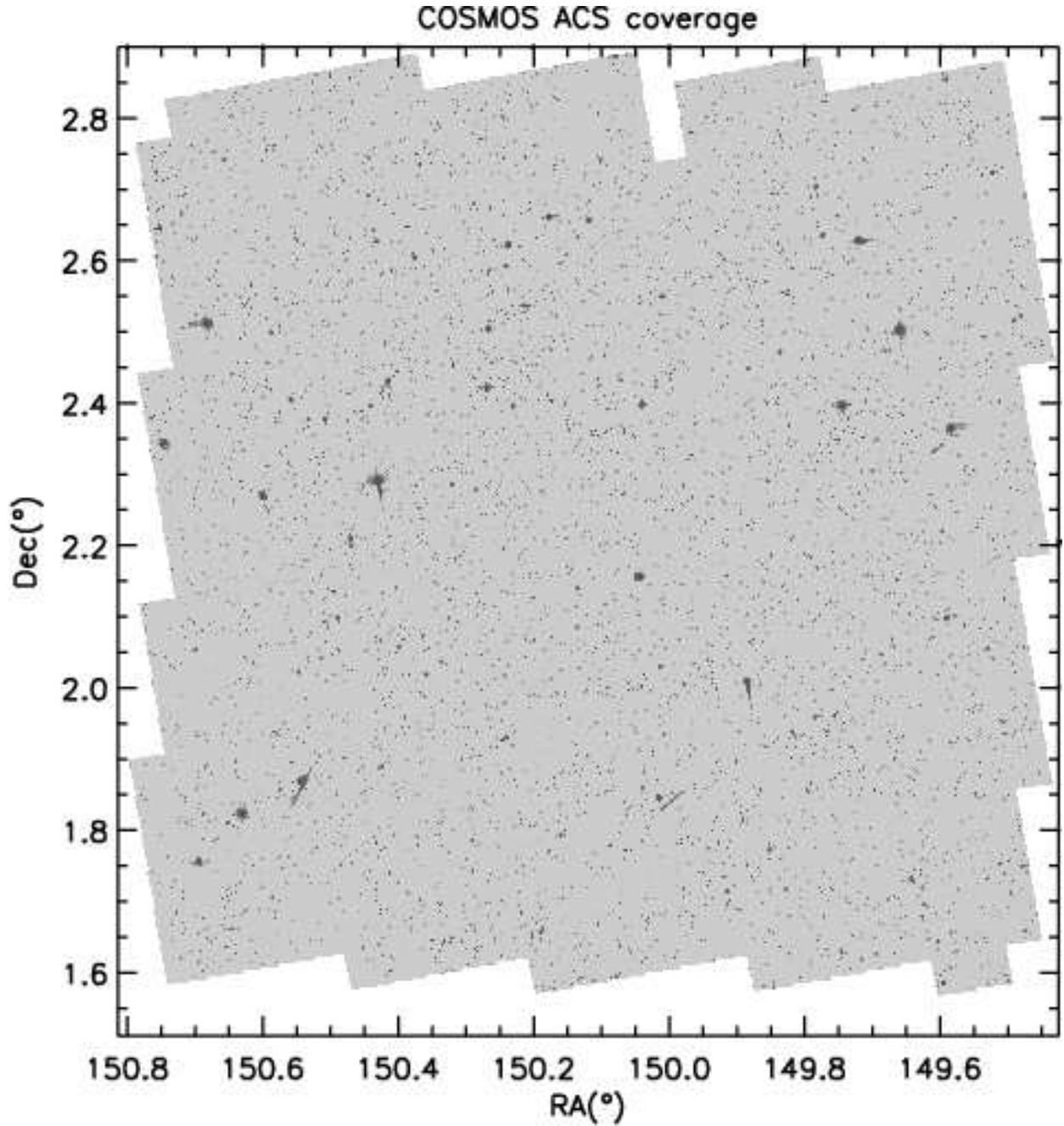}

\caption{The layout of the ACS mosaic of 581 I-band pointings is shown. The rectangle fully enclosing
all the ACS imaging has lower left and upper right corners (RA,DEC J2000) at  (150.7988\deg,1.5676\deg) and  (149.4305\deg, 2.8937\deg).
The  WFPC2 (3000 \& 4500\AA) and NICMOS (1.6$\mu$m) images cover  approximately 55\% and 6\% of the ACS area.} 
\label{acs}
\end{figure}

 The sensitivities and resolutions for the COSMOS-HST imaging are 
 summarized in Table \ref{tbl-2} and compared with those of 
 other HST surveys in Table \ref{tbl-3}. In order to facilitate direct comparison 
 of the different surveys, we have used the instrument exposure time calculators (ETCs) 
 provided by STScI, rather than published sensitivities of the surveys 
 which were derived with differing assumptions. For ACS and WFPC2, the flux is normalized in the V
 band; for 
NICMOS, the flux is normalized to 1.6$\mu$m. (The V-band normalization 
was adopted to be consistent with Table \ref{tbl-3} where the other HST surveys 
which cover several bands are summarized.) Normalization of the flux to 
I-band  reduces the ACS magnitudes by 1.4 mag (e.g. 28.6 $\rightarrow$ 27.2 mag).
The COSMOS ACS I-band coverage is shown in 
Figure  \ref{acs}. 
The rectangle bounding all the ACS imaging has lower left and 
upper right corners (RA,DEC J2000) at  
(150.7988\deg,1.5676\deg) and (149.4305\deg, 2.8937\deg). 
The positions observed in the NICMOS parallel observations are 
shown in  Figure  \ref{nicmos}; the locations of the WFPC2 parallels are not 
shown given in view of their low sensitivity (see below).

\section{Photometric Catalogs}\label{sec-cat}

The primary reference catalogs for COSMOS have objects selected from both the 
ACS images and the 
very deep, multi-band Subaru-SCAM COSMOS imaging  \citep{tan06}. The SCAM data 
 are of similar depth to the ACS imaging (approximately 0.8 mag deeper for 
 sources $> 1$\arcsec~ in diameter but for sources smaller than $\sim$
 0.3\arcsec, the ACS data are more sensitive). The ground-based imaging \citep{cap06} 
 presently includes 18 filters  including 
 narrow and intermediate bandwidth filters. Catalogs were made from both the ACS and Subaru I-band and NICMOS H-band imaging. The ACS 
 catalog comprises 1.2 million objects (\citep{lea06}; the NICMOS catalog 
 has 21639 objects ($> 1.5\sigma$ for 9 adjacent pixels).  For the ACS catalog, SExtractor \citep{ber96}
was used with the requirement of  $> 0.4\sigma$ in at least 4 adjacent pixels and the total isophotal 
signal-to-noise ratio $> 1$.
The ACS catalog includes additional internal sub-structures 
within sources listed in the lower-resolution, ground-based catalog. These catalogs 
and the derived photometric redshifts \citep{mob06} are presented and described in detail elsewhere \citep{koe06,cap06,lea06}.

\section{Public Data Release}

The COSMOS HST data are publicly available in staged releases (following calibration 
and validation) through the web sites for IPAC/IRSA : {\bf \url{http://irsa.ipac.caltech.edu/data\\/COSMOS/}} 
and STScI-MAST : {\bf \url{http://archive.stsci.edu/}}.~~
 (The STScI pipeline processed 
images are of course also available in the STScI archive.) The COSMOS ACS imaging is in the form
of separate drizzled images for each pointing, rotated and resampled to have North up 
with pixel scale 0.05\arcsec. IRSA also supplies a cutout capability derived 
from the full field mosaic (50 GB). The cutouts can be made with any field center and size;
and multiple cutouts are provided based on a user-supplied file containing source positions and field sizes. 
The SExtractor catalogs for ACS and NICMOS (see section \ref{sec-cat}) are also available through IRSA. 

\section{Source Counts and Completeness}

Figures  \ref{acs_counts} and  \ref{nicmos_counts} show the magnitude distributions of sources  in the ACS and NICMOS source catalogs
(Section \ref{sec-cat}). The ACS source counts are listed in Table \ref{tbl-4}. We have compared the ACS I-band source counts with those 
published by \cite{fer00} for HDF. In HDF, the derived I-band count at I$_{AB} = 25 $ mag is 
$1.0\times10^5$ galaxies per square degree per $\Delta m = 1$ mag. For the COSMOS ACS 
catalog the count is $\sim$21,000 for 1.8 square degrees per $\Delta m = 0.1$ mag (at I$_{AB} = 25 $ mag, see Figure \ref{acs_counts}). The COSMOS counts are $\sim 16$\% higher than those in the HDFs, 
but given the fact that different SExtraction parameters were probably used in COSMOS and HDF 
the agreement is quite acceptable.  The H-band integrated number counts are equivalent to
3478, 14130 and 41300 deg$^{-2}$ per 0.5 mag bin at H $=$ 20, 22 and 24 mag (AB). A recent compilation of previous 
surveys by \cite{fri05} has approximately 3300, 15000 and 41000 deg$^{-2}$ 
in the same bins (see dashed curve in their Figure 1).

The completeness of the ACS catalog was determined using the 
standard  technique of inserting false sources of specified half-light size and total flux 
\citep{gia04}. Size and flux must both be explored since the ability to detect sources 
depends on their surface brightness and hence their flux and size. The simulated galaxies were a 50/50 mix of exponential disks and r$^{1/4}$ spheroids \citep{gia04}. 
Figure  \ref{acs_completeness} shows contours for the percentage of the test galaxies recovered. For galaxies 
with half-light radii of 0.25, 0.5 and 1\arcsec, the completeness is $\sim$50\% at I$_{AB} \simeq$ 26.0, 24.7 and 24.5 mag. 

\section{Analysis Enabled by COSMOS ACS Observations}
 
The high resolution ACS imaging  is critical to the COSMOS survey, providing:
galaxy morphologies, multiplicities and merger rates out to z $\geq 2$, environmental density from DM maps at z $\leq 1$, and size and limited morphological information at redshifts out to z = 6. 
The morphological parameters 
obtained from the ACS imaging include bulge/disk ratios, 
concentration, asymmetry, size, multiplicity, color, clumpiness (see \cite{cas06,sca06,cap06}). 
The COSMOS I-band ACS images have sufficient 
depth and resolution to allow classical bulge-disk decomposition for 
{\it L$^*$} galaxies at $z\leq$ 2, while
less detailed structural parameters such as concentration, asymmetry, 
clumpiness and size can be measured for all galaxies down to the 
COSMOS spectroscopic survey limit (37,500 galaxies with I$_{AB}$ $\leq$25; \cite{lil06,imp06}), out to z $\sim 5$. COSMOS ACS imaging has been crucial for the identification and analysis
of galactic interactions and mergers (e.g. \cite{kar06}) -- processes which are central to 
the early evolution of galaxies.

For the purposes of weak lensing analysis, approximately 87 galaxies per $\sq$\arcmin  were sufficiently resolved  with ACS (c.f. $\sim30$ $\sq$\arcmin ~from Subaru-SCAM). Their median (mean) redshift is 1.02 (1.25), and their per-component rms shear is 0.309. This permits mass reconstructions with an optimal resolution on the sky at scales $\sim100\arcsec$, and a redshift sensitivity that peaks between $z=0.2$ and $z=0.6$. These parameters enable detection of  $\sim 7\times10^{13}M_\sun$ cluster at $z=0.2$ with $5\sigma$ signal-to-noise ratio \citep{rho06,mas06}.

 
 \acknowledgments
  The HST COSMOS Treasury program was supported through NASA grant
HST-GO-09822. We wish to thank Tony Roman, Denise Taylor, and David 
 Soderblom for their assistance in planning and scheduling of the extensive COSMOS 
 observations.
 We gratefully acknowledge the contributions of the entire COSMOS collaboration
 consisting of more than 70 scientists. 
 More information on the COSMOS survey is available \\ at
  {\bf \url{http://www.astro.caltech.edu/$\sim$cosmos}}. It is a pleasure the 
 acknowledge the excellent services provided by the NASA IPAC/IRSA 
 staff (Anastasia Laity, Anastasia Alexov, Bruce Berriman and John Good) 
 in providing online archive and server capabilities for the COSMOS datasets.
 The COSMOS Science meeting in May 2005 was supported in part by 
 the NSF through grant OISE-0456439. We thank Rob Kennicutt for suggestions 
 on the manuscript. 
 
 
 
 {\it Facilities:} \facility{HST (ACS)}, \facility{HST (NICMOS)}, \facility{HST (WFPC2)}.

\clearpage
 
\begin{deluxetable}{lccrccr}
 \tabletypesize{\scriptsize}
 \tablecaption{HST Observations\label{tbl-1}}
 \tablewidth{0pt}
 \tablehead{
 \colhead{Instrument} & \colhead{Filter} & \colhead{Mode\tablenotemark{a} } & \colhead{Orbits}  & \colhead{Exposure (sec) }  & \colhead{Dates}  & \colhead{Proposal ID}  } 
 \startdata
ACS & F814W (I-band)   & P & 261 & 2028 & 10/15/03 $\rightarrow$ 5/21/04  & 09822 (Cy 12) \\ 
ACS & F475W (g-band)   & P & 9 & 2028 & 10/15/03 $\rightarrow$ 5/21/04  & 09822 (Cy 12) \\ 
ACS & F814W (I-band)   & P & 320 & 2028 & 10/15/04 $\rightarrow$ 5/21/05  & 10092 (Cy 13) \\ 
NICMOS-NIC3 & F160W (H-band)   & PP & 225 & 1536 & 10/15/03 $\rightarrow$ 5/21/04  & 09999 (Cy 12) \\ 
NICMOS-NIC3 & F160W (H-band)   & PP & 282 & 1536 & 10/15/04 $\rightarrow$ 5/21/05  & 10337 (Cy 13) \\ 
WFPC2 & F300W (U-band)   & CP & 270 & 1600 & 10/15/03 $\rightarrow$ 5/21/04  & 09822 (Cy 12) \\ 
WFPC2 & F300W (U-band)   & CP & 149 & 1600 & 10/15/04 $\rightarrow$11/21/04  & 10092 (Cy 13) \\ 
WFPC2 & F450W (B-band)   & CP & 171 & 1600 & 11/21/04 $\rightarrow$ 5/21/05  & 10092 (Cy 13) \\ 
 \enddata
 \tablenotetext{a}{Scheduling Mode : P -- primary, PP -- pure-parallel, CP -- coordinated-parallel.}
  \end{deluxetable}

\begin{deluxetable}{lcccccc}
 \tabletypesize{\scriptsize}
 \tablecaption{COSMOS HST Sensitiivities and Resolution\label{tbl-2}}
 \tablewidth{0pt}
 \tablehead{
 \colhead{Instrument} & \colhead{Filter} &  \colhead{$5\sigma_{AB}$\tablenotemark{a} } & \colhead{$5\sigma_{STmag}$} & \colhead{$5\sigma_{Vega}$} & \colhead{Res(FWHM)} & \colhead{Pixel} } 
 \startdata
ACS & F814W (I-band)   & 28.6 & 29.5 & 28.2 & 0.09\arcsec & 0.05\arcsec \\ 
ACS & F475W (g-band)   & 27.9 & 27.6 & 28.0  & 0.05\arcsec & 0.05\arcsec \\ 
NICMOS-NIC3 & F160W (H-band)   & 25.9 & 28.3 & 24.6  & 0.16\arcsec & 0.20\arcsec\\ 
WFPC2 & F300W (U-band)   & 24.8 & 26.1 & 24.8  & 0.10\arcsec & 0.1(0.046,PC)\arcsec\\ 
WFPC2 & F450W (B-band)   & 26.7 & 26.3 & 26.8  & 0.10\arcsec & 0.1(0.046,PC)\arcsec\\ 
 \enddata
 \tablenotetext{a}{Sensitivities for optimally-extracted point sources with a
$\lambda^{-1}$ power law spectrum for ACS and NICMOS. For WFPC2, assumes
an A0 star spectrum. For ACS and WFPC2, the flux is normalized to V, for 
NICMOS the flux is normalized to 1.6$\mu$m. The V-band normalization 
was adopted to be consistent with Table \ref{tbl-3} where surveys in several bands are compared. Normalization of the flux to 
I-band reduces the ACS magnitudes by 1.4 mag (e.g. 28.6 $\rightarrow$ 27.2).
}
  \end{deluxetable}

 \begin{deluxetable}{lccccc}
 \tabletypesize{\scriptsize}
 \tablecaption{Relative Pt. Source Sensitivities\label{tbl-3}}
 \tablewidth{0pt}
 \tablehead{
 \colhead{Filter} & \colhead{Survey}  & \colhead{$5\sigma$ AB\tablenotemark{a} } &
 \colhead{Orbits} & \colhead{Vega - AB} & \colhead{STmag - AB}} 
 \startdata
F435 & UDF   & 29.94 & 56 & 0.11 &  0.52 \\
~ ..    & GOODS & 28.32 & 3 & ... &   ... \\
F606     & UDF   & 30.84 & 56 & 0.09 & 0.17\\
~ .. & GOODS & 29.14 & 2.5 & ... & ... \\
F775     & UDF   & 31.30 & 150 & -0.40 &  0.74 \\
~ .. & GOODS & 29.04 & 2.5 & ... & ... \\
{\bf F814} & {\bf COSMOS }  & {\bf 28.63} & {\bf 1} & {\bf -0.44} & {\bf 0.84} \\
F850 & GOODS &  29.11 & 5 & -0.54 &  1.09 \\
~ ..     & UDF &  30.99 & 150 & ... & ... \\


 \enddata
 %
\tablecomments{HDF : \cite{wil96,wil00}; GOODS : \cite{gia04} ; UDF : \cite{bec06} .
To facilitate direct comparison 
 of the different surveys, we have used the instrument exposure time calculators (ETCs) 
 provided by STScI, rather than published sensitivities of the surveys 
 (which were derived with differing assumptions).}

 \tablenotetext{a}{Sensitivities for optimally-extracted point sources;
for a source uniformly extended over $\sim$0.25\arcsec ~diameter, the
limiting magnitudes 
are $\sim$1 mag greater. (Assumes: 2028s per orbit for COSMOS and 2500s per orbit for the other surveys,
a $\lambda^{-1}$ power law spectrum with normalization at V, 4 cr-split/orbit for COSMOS, 2 cr-split/orbit for GOODS and UDF, no reddening)}
 \end{deluxetable}

 \begin{deluxetable}{lrr}
 \tabletypesize{\scriptsize}
 \tablecaption{ACS Source Counts\label{tbl-4}}
 \tablewidth{0pt}
 \tablehead{
 \colhead{I $< $ mag} & \colhead{all objects}  & \colhead{non-stellar} } 
 \startdata
25 & 288,657   & 266,039 \\
26   & 567,143 & 531,982 \\
27     & 1,029,007   & 878,445 \\
 \enddata
\tablecomments{Source counts in F814W ACS images obtained using SExtractor as described in text.
 The ACS catalog used for lensing studies is present in \citep{lea06}. }

 \end{deluxetable}
 
 \clearpage
 
 
 
\begin{figure}[ht]
\epsscale{1.0} 
\includegraphics[angle=90]{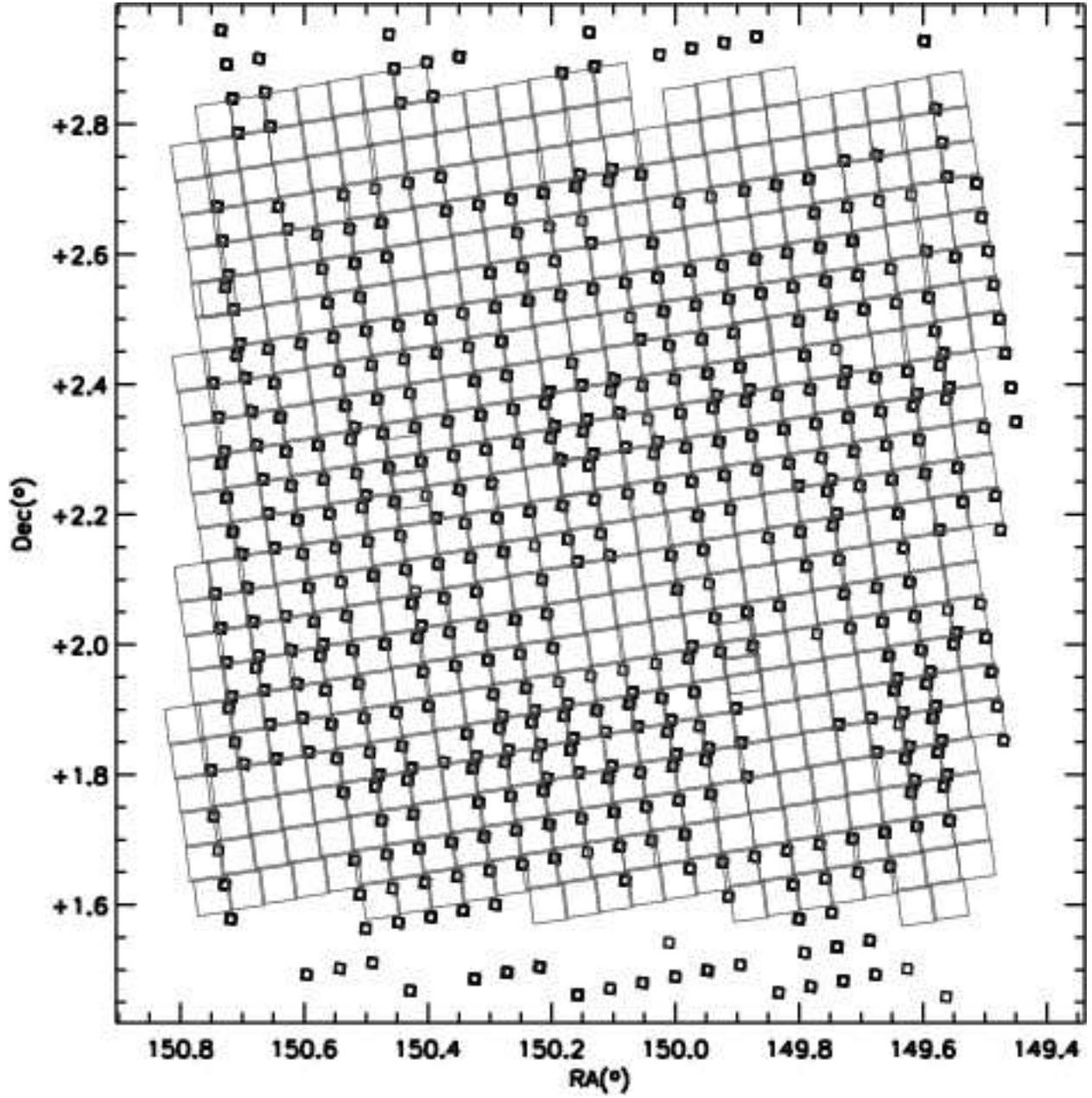}
\caption{The layout of the NICMOS parallels is shown superposed on the ACS pointings.
The NICMOS (1.6$\mu$m) images cover  approximately 6\% of the ACS area.} 
\label{nicmos}
\end{figure}

\begin{figure}[ht]
\epsscale{1.0}
\plotone{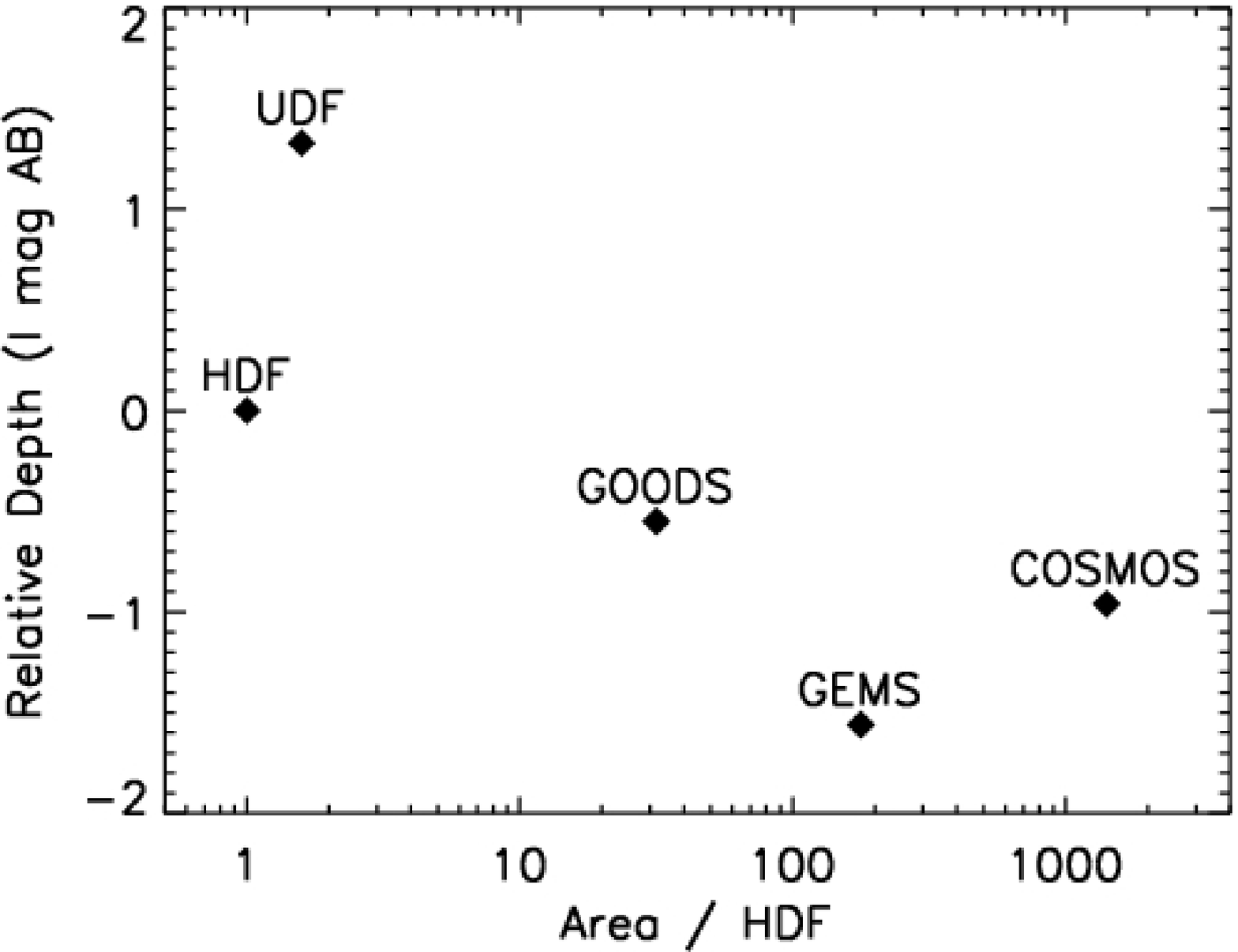}
\caption{Relative field areas and sensitivities of major
HST 
surveys at $\sim$8000\AA 
~compared to the original HDF survey. COSMOS has 9$\times$ the area 
of GEMS (\cite{rix04}, the next largest survey) with sensitivity just 1.4$\times$
less 
than GOODS (\cite{gia04}, in 20\% the time due to the higher throughput
of F814W vs F850LP -- GOODS). The relative sensitivities shown here 
were derived using the instrument exposure time calculators (ETCs for an optimally extracted 
point source) in order to 
facilitate equivalent comparisons.
} 
\label{surveys}
\end{figure}

\begin{figure}[ht]
\epsscale{1.} 
\plottwo{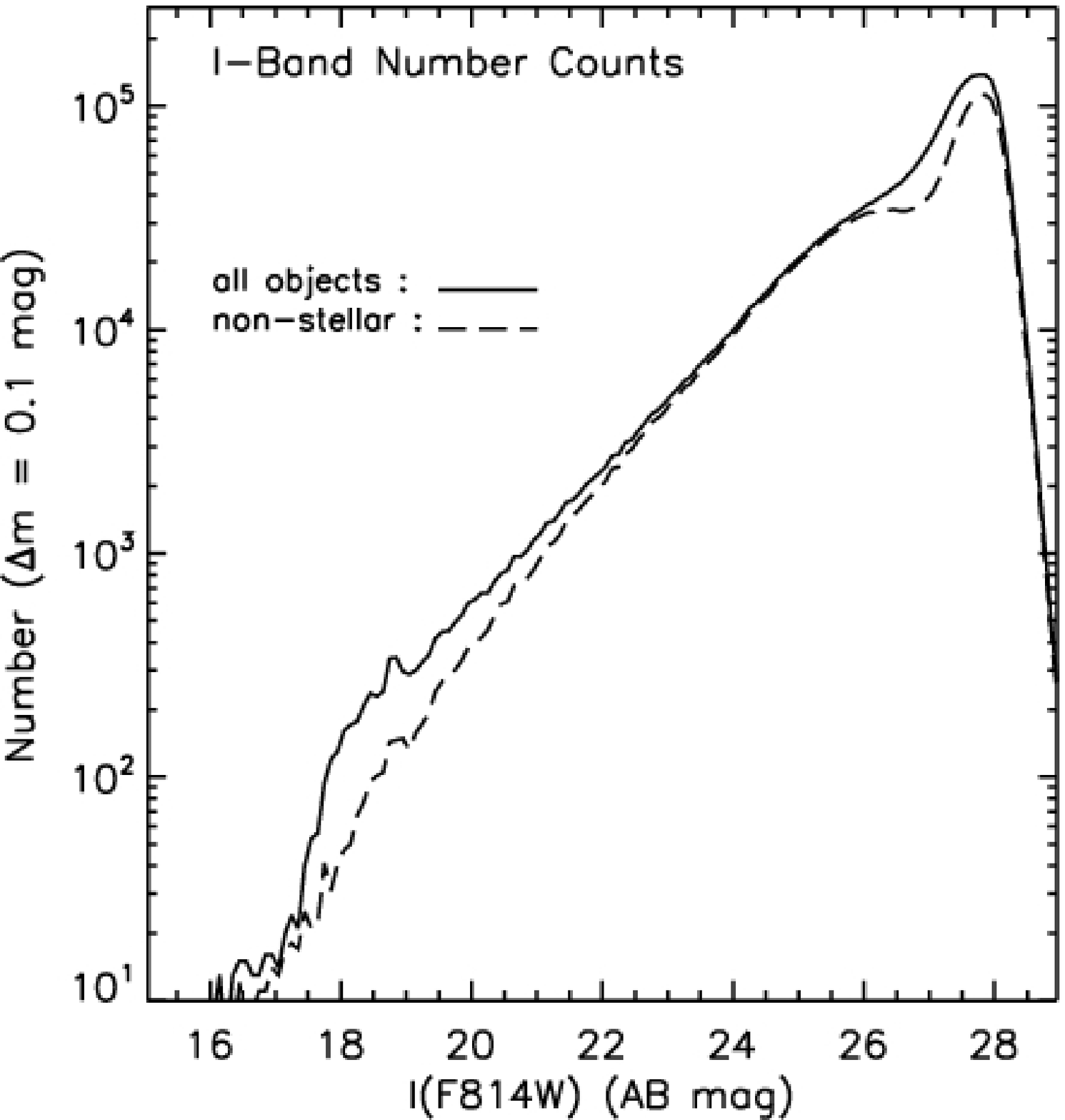}{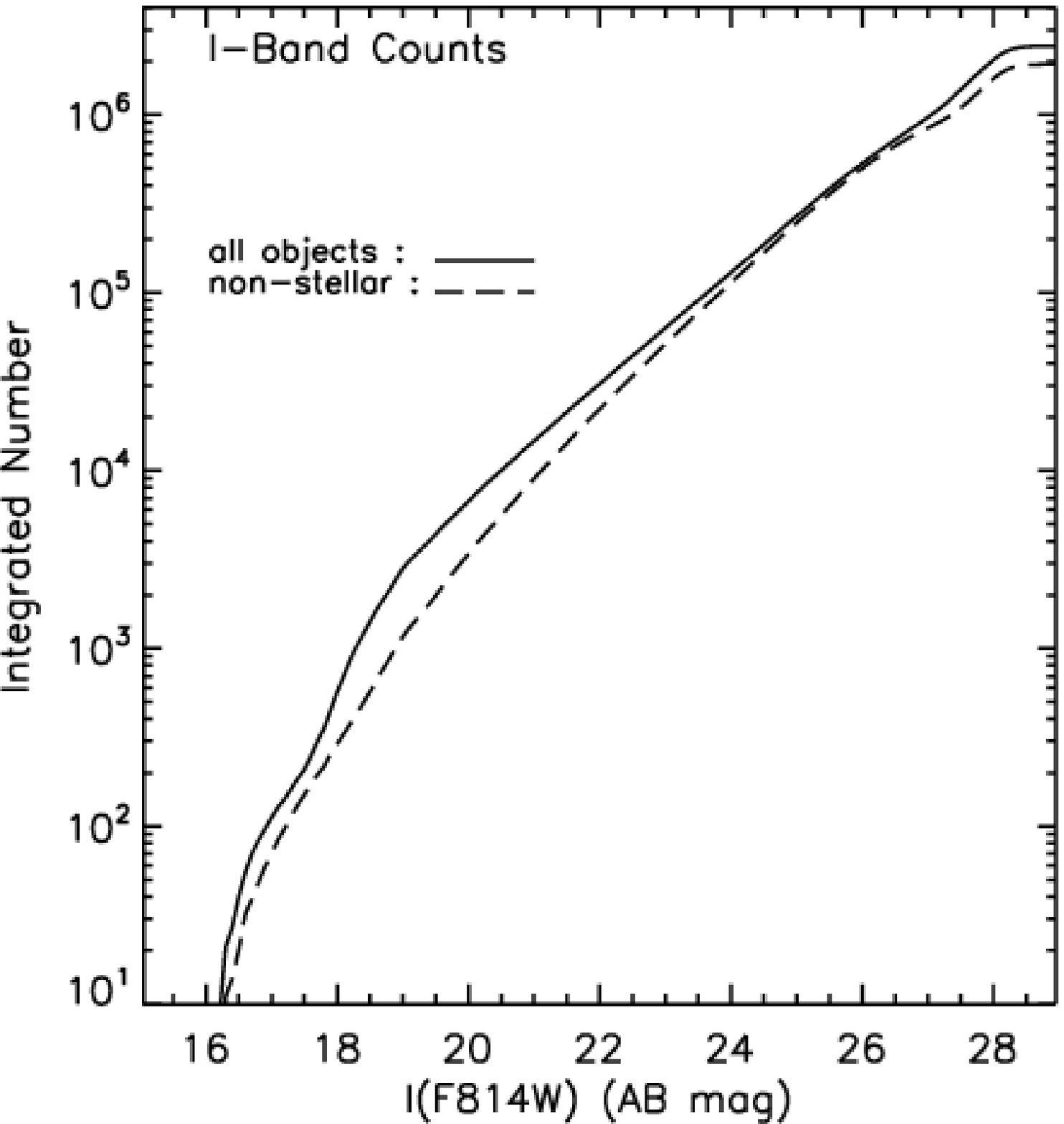}
\caption{Source counts (differential -- top panel ; integral -- lower panel) for the COSMOS ACS I-band catalog (from SExtractor
requiring $> 0.4\sigma$ in at least 4 adjacent pixels and the total isophotal 
signal-to-noise ratio $> 1$.). The number counts for auto-mags are shown for 
for all objects (solid line) and those with 'stellarity' $< 0.95$ (dashed line). The upturn at 
I $< 27$ mag is the result of low SNR spurious detections. These counts are not corrected for 'completeness'.}
\label{acs_counts}
\end{figure}

\begin{figure}[ht]
\epsscale{1.} 
\plottwo{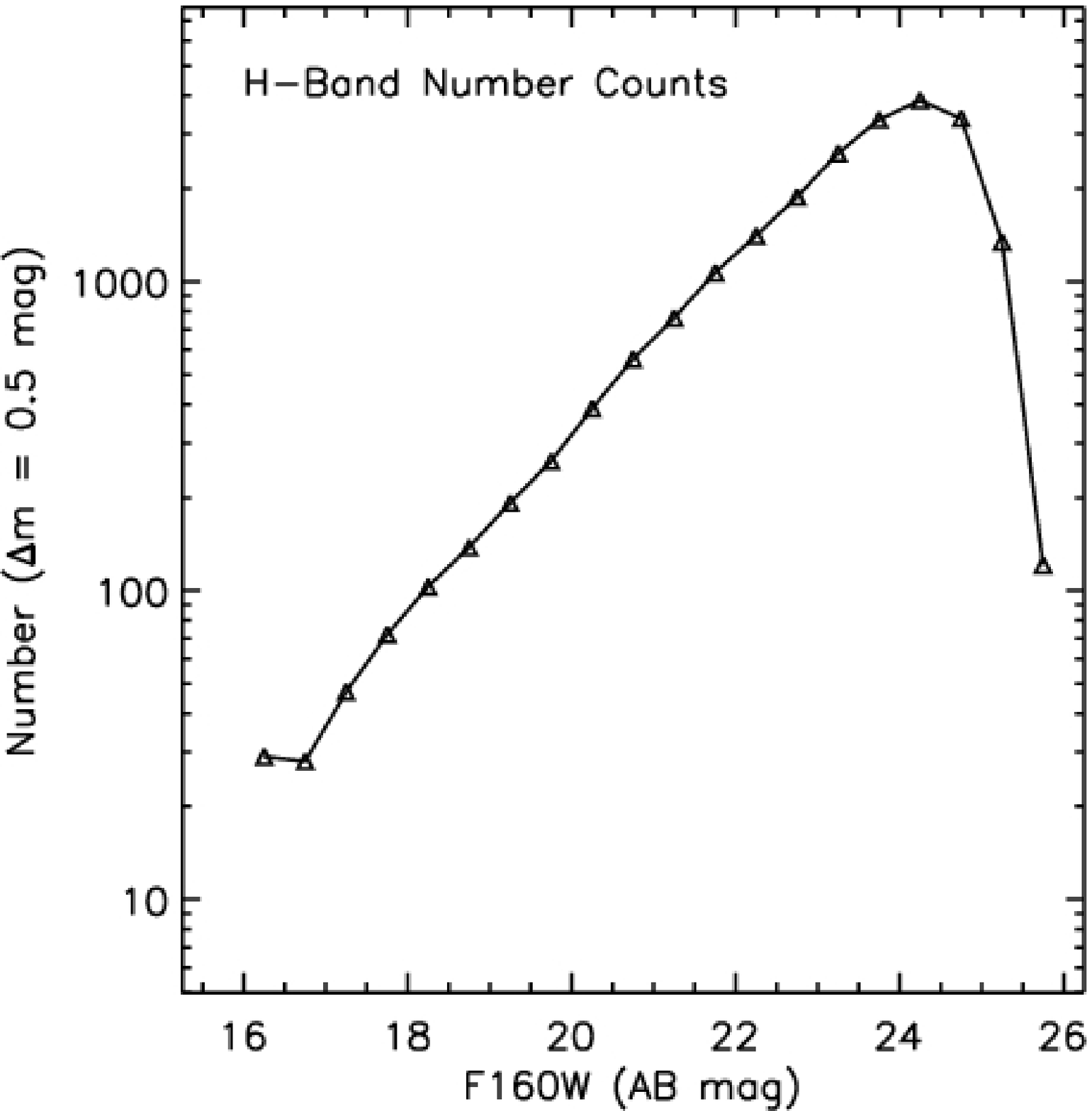}{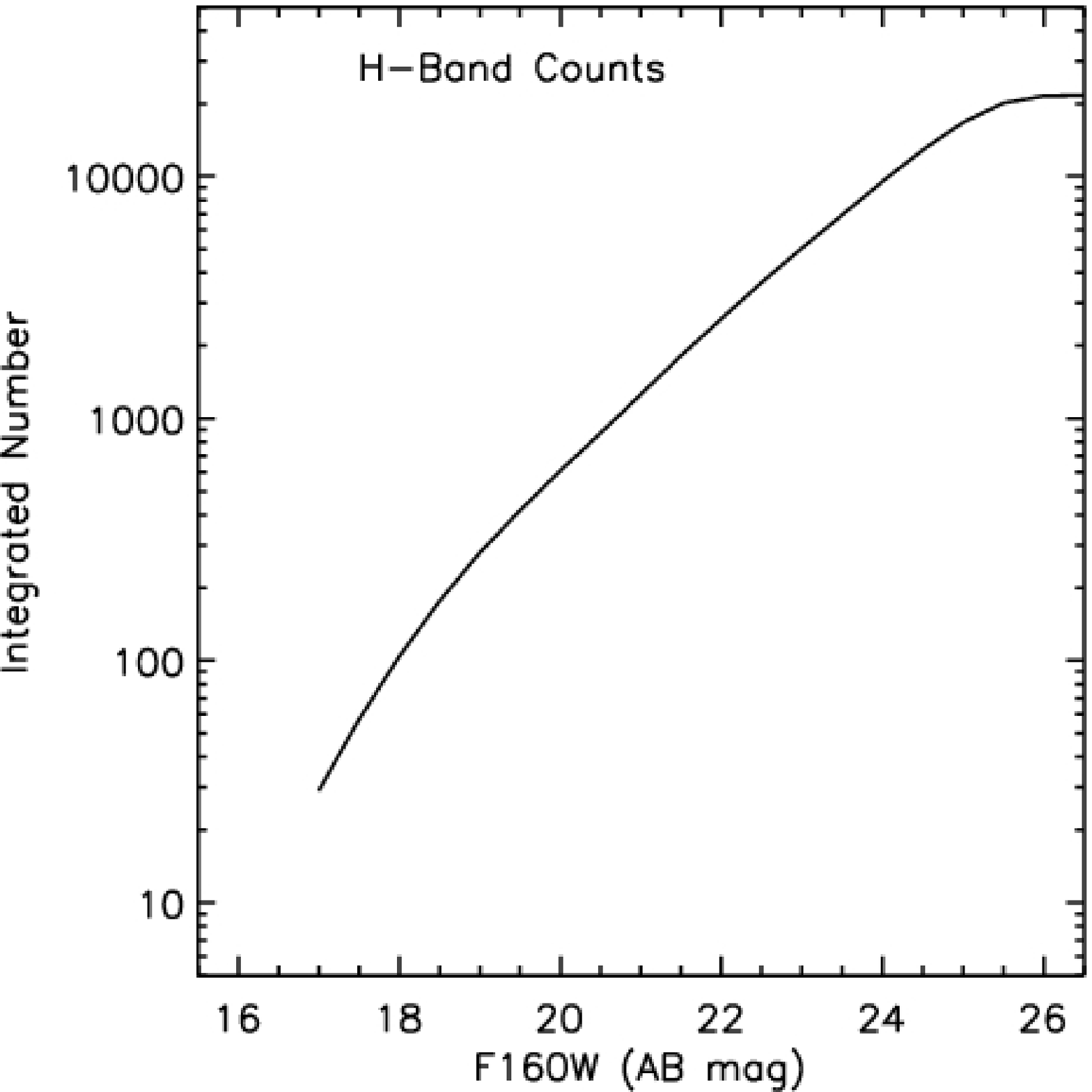}
\caption{Source counts (differential -- top panel ; integral -- lower panel) for the COSMOS NICMOS H-band catalog (based on SExtractor measurements requiring $> 1.5\sigma$ signal in 9 adjacent pixels). The number counts are for 0.5 mag bins; 
the total area imaged in NICMOS is  0.099 \sq\deg. The total number of objects is 21639. }
\label{nicmos_counts}
\end{figure}

\begin{figure}[ht]
\epsscale{1.0} 
\includegraphics[angle=90]{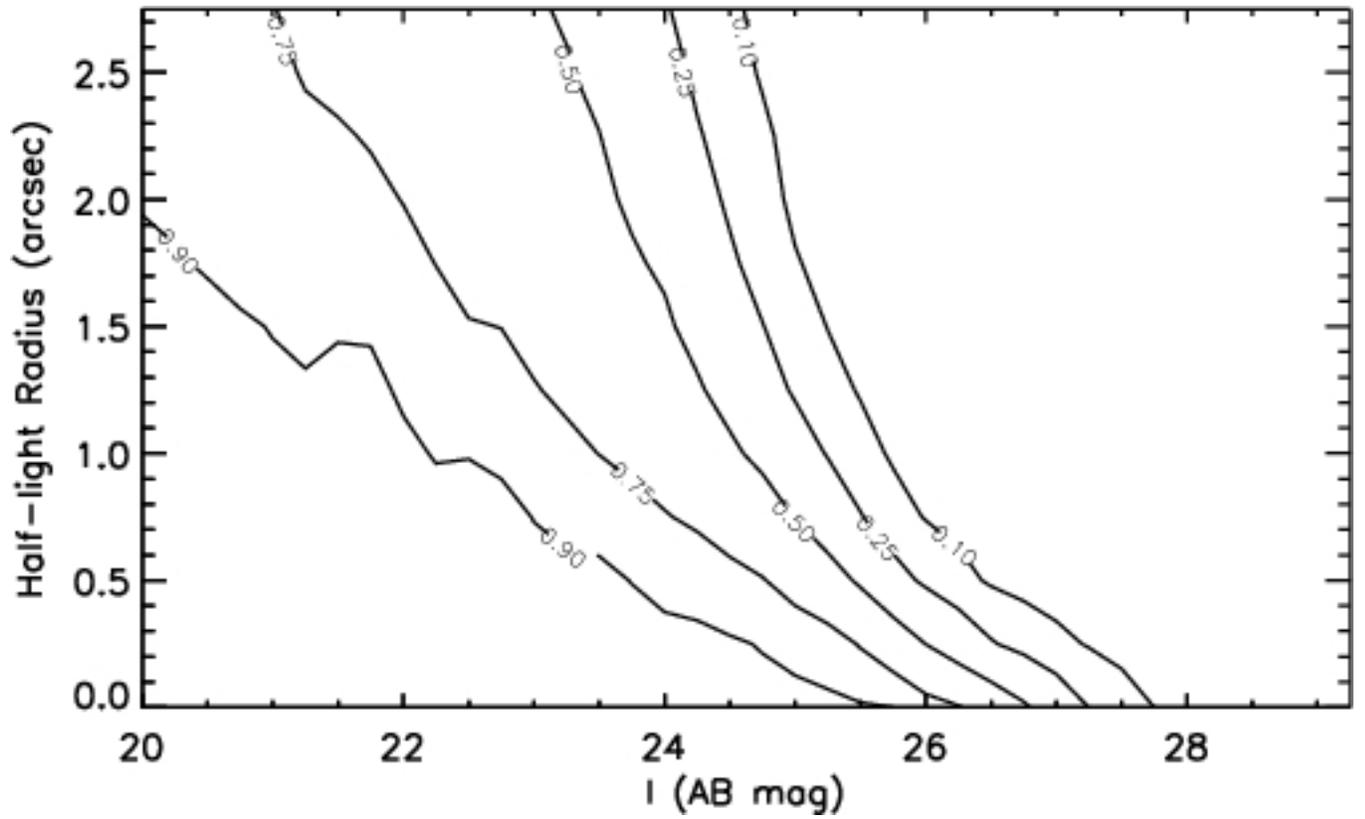}
\caption{Completeness estimates are shown for COSMOS ACS I-band imaging. Contours show the 
percentage of simulated sources (a 50/50 mix of exponential disks and spheroids) which were recovered by SExtractor as a function of sources
total magnitude and half-light radius.}
\label{acs_completeness}
\end{figure}

 
 \end{document}